\documentclass{article}%
\usepackage{amsmath}
\usepackage{amsfonts}
\usepackage{amssymb}
\usepackage{graphicx}%
\providecommand{\U}[1]{\protect\rule{.1in}{.1in}}

\begin{document}
\title{\textbf{Conserved current for the Cotton tensor,
black hole entropy and equivariant Pontryagin forms}}

\author{Roberto Ferreiro P\'{e}rez\\
Departamento de Econom\'{\i}a Financiera
y Contabilidad I, UCM.\\
Campus de Somosaguas, 28223-Pozuelo de Alarc\'on,
Spain.\\
\emph{E-mail:\/} \texttt{roferreiro@ccee.ucm.es}}
\date{}

\maketitle

\begin{abstract}
The Chern-Simons lagrangian density in the space of metrics of a
$3$-dimensional manifold $M$ is not invariant under the action of
diffeomorphisms on $M$. However, its Euler-Lagrange operator can
be identified with the Cotton tensor, which is invariant under
diffeomorphims. As the lagrangian is not invariant, Noether
Theorem cannot be applied to obtain conserved currents. We show
that it is possible to obtain an equivariant conserved current for
the Cotton tensor by using the first equivariant Pontryagin form
on the bundle of metrics. Finally we define a hamiltonian current
which gives the contribution of the Chern-Simons term to the black
hole entropy, energy and angular momentum.

\end{abstract}



\maketitle

\section{Introduction}
In Topologically Massive Gravity the lagrangian is given by the
Hilbert Einstein lagrangian plus a Chern-Simons term (e.g. see
\cite{TMG}). In dimension $3$, although the Chern-Simons
lagrangian for metrics is not invariant under the action of the
group of diffeomorphisms on the manifold, its the Euler-Lagrange
operator can be identified with the Cotton tensor which is
invariant. In fact, the Cotton tensor does not admit a
diffemorphisms invariant lagrangian (see \cite{anderson} and
references therein for the properties and history of the Cotton
tensor).

Since the Chern-Simons lagrangian is not diffeomorphisms
invariant, we cannot apply Noether theorem in order to obtain the
corresponding conserved currents. This note aims to show that it
is possible to define an equivariant conserved current. We show in
Section \ref{current} that this current appears in a natural way
from the geometry of the jet bundle of metrics and that it is
provided by the equivariant Pontryagin forms defined in \cite{WP}.
As the conserved currents associated to invariance under
diffeomorphisms are globally exact on shell we have $J_{CS}\approx
dQ_{CS}(X)$, where $Q_{CS}$ is the Noether charge. We show that
$Q_{CS}$ is given by the Schouten tensor of the metric. Finally in
Section \ref{entropy} we follow Wald's Noether method to compute
the contribution of the Chern-Simons term to the black hole
entropy. As the Chern-Simons lagrangian is not invariant, the
current $Q_{CS}$ does not give the correct value of the entropy.
We define a Hamiltonian current $q_{CS}$ for the Chern-Simons term
 by adding to the Noether charge $Q_{CS}$ an
additional term which is also obtained from the equivariant
Pontryagin form. We show that this hamiltonian current coincides
with a current defined in \cite{MO} for constant vector fields,
and hence gives the same value for the contribution of the
Chern-Simons term to the black-hole energy and angular momentum.
Moreover, we show that the current $q_{CS}$ also gives the correct
value of the contribution of the Chern-Simons term to the
black-hole entropy computed in \cite{Tach}.

We use the approach to the calculus of variations in terms of
differential forms on jet bundles. In Section \ref{jet} we recall
the basics results on the geometry of the jet bundle of the bundle
of metrics $J\mathcal{M}_{M}$, and we define the Pontryagin forms,
equivariant Pontryagin forms and Chern-Simons forms on
$J\mathcal{M}_{M}$.

Most of our results are based on very general properties of the
geometry of the jet bundle that can be easily generalized to
higher dimensions. However the final result for the black hole
entropy is not so easily obtained in higher dimensions and for
this reason we consider only the $3$-dimensional case.

In the following, the word metric means Riemannian or
pseudo-Riemannian metric.

\section{The Cotton tensor and the Chern-Simons lagrangian\label{CS}}

Let us recall the computation\ of the Euler Lagrange operator of
the gravitational Chern-Simons lagrangian. We follow the
exposition in \cite{guralnik}.

In dimension $3$ the Chern-Simons lagrangian for metrics on a
$3$-manifold $M$ is given locally by
\begin{equation}
\lambda_{CS}=\alpha\mathrm{tr}\left(  \Gamma\wedge
d\Gamma+\frac{2} {3}\Gamma\wedge\Gamma\wedge\Gamma\right)  ,
\label{csl}
\end{equation}
(in Section \ref{jet} we give a definition of Chern-Simons
lagrangian valid globally).

If we consider an arbitrary variation of the metric $\delta g_{ab}
$, then the variation of $\lambda_{CS}$ is given by
\begin{equation}
\delta\lambda_{CS}=2\alpha\mathrm{tr}\left( \delta\Gamma\wedge
R\right)  +\alpha d\left(  \delta\Gamma\wedge\Gamma\right)
\label{Dlambda}
\end{equation}

Using the expression for the variation of the Christoffel symbols
\[
\delta\Gamma_{bc}^{a}=\frac{1}{2}g^{aj}\left(  \nabla_{b}\delta
g_{jc} +\nabla_{c}\delta g_{jb}-\nabla_{j}\delta g_{bc}\right)
\]
and the expression for the Riemann tensor in terms of the Ricci
tensor in dimension $3$
\begin{equation}
R_{bcd}^{a}= \delta_{c}^{a}\left( R_{bd}-\frac{1}{2}g_{bd}R\right)
-\delta_{d}^{a}\left( R_{bc}-\frac{1}{2}g_{bc}R\right)
+g_{bd}R_{c} ^{a} -g_{bc}R_{d}^{a} \label{expR}
\end{equation}
we obtain
\[
\mathrm{tr}\left(  \delta\Gamma\wedge R\right) =\nabla_{b}\delta
g_{ia}R_{c}^{i}dx^{a}\wedge dx^{b}\wedge dx^{c}
\]

If we integrate by parts it follows that
\begin{eqnarray*}
\mathrm{tr}\left(  \delta\Gamma\wedge R\right)   &  =\left(
\partial_{b}\left(  \delta g_{ia}R_{c}^{i}\right)  -\delta g_{ia}\left(
\nabla_{b}R_{c}^{i}\right)  \right)  dx^{a}\wedge dx^{b}\wedge dx^{c}\\
&  =-\left(  d\left(  \delta g_{ia}R_{c}^{i}dx^{a}\wedge
dx^{c}\right) +\delta g_{ia}\left(  \nabla_{b}R_{c}^{i}\right)
dx^{a}\wedge dx^{b}\wedge dx^{c}\right)  .
\end{eqnarray*}

The first term is an exact form, and the second one can be
expressed in terms of the Cotton tensor
\[
C^{ab}=-\frac{1}{2\sqrt{-\left\vert g\right\vert }}\left(
\varepsilon
^{ija}\nabla_{i}R_{j}^{b}+\varepsilon^{ijb}\nabla_{i}R_{j}^{a}\right)
\]
and we obtain
\begin{equation}
\mathrm{tr}\left(  \delta\Gamma\wedge R\right)  =\delta g_{ij}
C^{ij}\mathrm{vol}+dN(\delta g) \label{EP1}
\end{equation}
where $\mathrm{vol}=\sqrt{\left\vert g\right\vert }d^{3}x$ and
$N(\delta g)=-\delta g_{ia}R_{c}^{i}dx^{a}\wedge dx^{c}$.

By replacing (\ref{EP1}) on (\ref{Dlambda})\ we obtain the first
variational formula

\begin{equation}
\delta\lambda_{CS}=2\alpha\delta g_{ij}C^{ij}\mathrm{vol}+\alpha
d(N(\delta g)+\mathrm{tr}(\delta\Gamma\wedge\Gamma))
\label{dlambda}
\end{equation}
Hence the Euler-Lagrange operator of $\lambda_{CS}$ can be
identified with the Cotton tensor.

Now we consider the natural action of the diffeomorphism group of
$M$ in the space of metrics on $M$. Let us consider a variation of
the metric $\delta _{X}g=-L_{X}g$ induced by an infinitesimal
diffeomorphism $X\in\mathfrak{X} (M)$. If in local coordinates we
have $X=X^{i}\partial/\partial x^{i}$ then
\begin{equation}
\delta_{X}g_{ab}=-\left(
\partial_{k}g_{ab}X^{k}+g_{kb}\partial_{a}
X^{k}+g_{ak}\partial_{b}X^{k}\right)  =-\left(
g_{ja}\nabla_{b}X^{j} +g_{jb}\nabla_{a}X^{j}\right)  \label{expdg}
\end{equation}

By replacing (\ref{expdg}) on (\ref{dlambda}) we obtain
\begin{equation}
\delta_{X}\lambda_{CS}=2\alpha\delta_{X}g_{ab}C^{ab}\mathrm{vol}+\alpha
d(N(\delta_{X}g)+\mathrm{tr}(\delta_{X}\Gamma\wedge\Gamma)).
\label{LVF}
\end{equation}

It can be seen (see Section \ref{CSF}) that we have
$\delta_{X}\lambda_{CS}=d\sigma$ for certain form $\sigma$, and
hence
$J_{CS}=N(\delta_{X}g)+\mathrm{tr}(\delta_{X}\Gamma\wedge\Gamma)-\sigma$
is a conserved current, i.e.
$dJ_{CS}=2\alpha\delta_{X}g_{ab}C^{ab}\mathrm{vol}\approx 0$. We
show in Section \ref{current} that this conserved current can be
obtained directly by combining equation (\ref{EP1}) with the
equivariant Pontryagin forms.

\section{Equivariant Pontryagin forms in the jet bundle of metrics\label{jet}}

\subsection{The jet bundle of the bundle of metrics}

Let us recall some results on the geometry of the jet bundle of
the bundle of metrics and the relation with the concepts
introduced in the previous section. We refer to
\cite{olver,saunders} for more details on the description of
variational calculus in terms of the geometry of jet bundles and
to (\cite{natconn,WP}) for the geometry of the bundle of metrics.

We denote by $J\mathcal{M}_{M}$ the jet bundle of the bundle of
metrics i.e., $J\mathcal{M}_{M}$ is the space obtained when the
derivatives of the metric are considered as independent variables.
Hence if $(x^{i})$ are coordinates on $M$, then the coordinates on
$J\mathcal{M}_{M}$ are $(x^{i},g_{ij}
,g_{ij,k},g_{ij,kr},\ldots)$.

Sometimes it is interesting to consider $J\mathcal{M}_{M}$ as an
infinitesimal version of the product $M\times\mathrm{Met}M$, where
$\mathrm{Met}M$ is the space of metrics on $M$. Both spaces are
related by the evaluation map $M\times\mathrm{Met}M\rightarrow
J\mathcal{M}_{M}$, $(x,g)\mapsto (x^{i},g_{ij}(x) ,\partial_k
g_{ij}(x),\ldots)$. For example, we have a canonical decomposition
$T^{\ast}J\mathcal{M}_{M}\cong T^{\ast}M\oplus
V^{\ast}(J\mathcal{M}_{M})$, where $V^{\ast}(J\mathcal{M} _{M})$
is the space contact (or vertical) $1$-forms generated by the
1-forms $\delta g_{ij,I}=Dg_{ij,I}-g_{ij,I+k}Dx^{k}$, and $D$
denotes the exterior differential on $\Omega^{r}
(J\mathcal{M}_{M})$. Accordingly we have
\begin{equation}
\Omega^{r}(J\mathcal{M}_{M})=\oplus_{p+q=r}\Omega^{p,q}(J\mathcal{M}_{M}),
\label{des}
\end{equation}
where $\Omega^{p,q}(J\mathcal{M}_{M})$ is the space of
$p$-horizontal and $q$-vertical forms. The exterior differential
$D$\ on $\Omega^{r} (J\mathcal{M}_{M})$ splits into horizontal and
vertical differentials $D=d+\delta$, where the horizontal
differential $d\colon\Omega^{p,q}
(J\mathcal{M}_{M})\rightarrow\Omega^{p+1,q}(J\mathcal{M}_{M})$
measures the changes on $M,$ whereas the vertical differential
$\delta\colon\Omega
^{p,q}(J\mathcal{M}_{M})\rightarrow\Omega^{p,q+1}(J\mathcal{M}_{M})$
measures the changes under variations of the metric. As a
consequence of $D^{2}=0$ we obtain
$d^{2}=\delta^{2}=d\delta+\delta d=0$. For example we have
$dx^{k}=Dx^{k}$, $\delta x^{k}=0$, $\delta
g_{ij,I}=Dg_{ij,I}-g_{ij,I+k} Dx^{k}$,
$dg_{ij,I}=g_{ij,I+k}Dx^{k}$.

The diffeomorphisms group of $M$ acts in a natural way on the
metrics on $M$ and induces an action on its derivatives. Hence
$\mathrm{Diff}M$ acts on $J\mathcal{M}_{M}$. At the infinitesimal
level, for every $X\in\mathfrak{X} (M)$ we obtain a vector field
$X_{J}\in\mathfrak{X}(J\mathcal{M}_{M})$.

Accordingly to the splitting (\ref{des}) the vector field $X_{J}$
can be expressed as $X_{J}=H_{X}+V_{X}$ where $H_{X}$ and $V_{X}$
are the horizontal and vertical components respectively. If in
local coordinates $X=X^{i}
\partial/\partial x^{i}$ then we have
\begin{eqnarray}
H_{X}  &  =X^{i}\frac{d}{dx^{i}},\label{HX}\\
V_{X}  &  =\delta_{X}g_{ij}\frac{\partial}{\partial
g_{ij}}+\frac{d(\delta
_{X}g_{ij})}{dx^{k}}\frac{\partial}{\partial
g_{ij,k}}+\ldots=\sum_{ij,I}
\frac{d^{|I|}(\delta_{X}g_{ij})}{dx^{I}}\frac{\partial}{\partial
g_{ij,I}}, \label{VX}
\end{eqnarray}
where the total derivatives are defined by
$\frac{d}{dx^{k}}=\frac{\partial }{\partial
x^{k}}+\sum_{ij,I}g_{ij,I+k}\frac{\partial}{\partial g_{ij,I}}$
and
\[
\delta_{X}g_{ij}=-\left(
g_{ij,k}X^{k}+g_{kj}\partial_{i}X^{k}+g_{ik}
\partial_{j}X^{k}\right).
\]
Note that the last expression is similar formula (\ref{expdg}) of
the previous section and that
$\delta_{X}g_{ij}=\iota_{V_{X}}\delta g_{ij}.$

If $\alpha\in\Omega^{p,q}(J\mathcal{M}_{M})$ then
$L_{X_{J}}\alpha\in \Omega^{p,q}(J\mathcal{M}_{M})$ and hence we
have
\begin{eqnarray*}
 L_{X_{J}}\alpha=\iota_{H_{X}}d\alpha+d\iota_{H_{X}}\alpha+\iota_{V_{X}}\delta\alpha
+\delta\iota_{V_{X}}\alpha ,\\
\iota_{H_{X}}\delta\alpha+\delta\iota_{H_{X}}\alpha    =0,\\
\iota_{V_{X}}d\alpha+d\iota_{V_{X}}\alpha   =0.
\end{eqnarray*}

The usual constructions in the calculus of variations can be
expressed as differential forms in $J\mathcal{M}_{M}$. For
example, if $\lambda$ is a lagrangian density then
$\lambda\in\Omega^{n,0}(J\mathcal{M}_{M})$ and
$\delta\lambda\in\Omega^{n,1}(J\mathcal{M}_{M})$.

We denote by
$F^{1}(J\mathcal{M}_{M})\subset\Omega^{n,1}(J\mathcal{M}_{M})$ the
subspace of forms of the type $A^{ij}\delta
g_{ij}\wedge\mathrm{vol}$ (usually this space it is called the
space of functional $1$-forms or source forms).  General results
on the variational bicomplex assert that for every $\alpha
\in\Omega^{n,1}(J\mathcal{M}_{M})$ we have
$\alpha=\mathcal{F}-d\theta$, where $\mathcal{F}\in
F^{1}(J\mathcal{M}_{M})$ and $\theta\in\Omega^{n-1,1}
(J\mathcal{M}_{M})$, and the form $\mathcal{F}$ is uniquely
determined by $\alpha$. Moreover, if $\alpha$ is
$\mathrm{Diff}M$-invariant then $\mathcal{F}$ and $\theta$ can be
chosen $\mathrm{Diff}M$-invariant.

In particular, if $\lambda\in\Omega^{n,0}(J\mathcal{M}_{M})$ is a
lagrangian density then we obtain the first variational formula
$\delta\lambda =\mathcal{E}-d\theta$, where
$\mathcal{E}=\mathcal{E}^{ij}\delta g_{ij} \wedge\mathrm{vol}\in
F^{1}$ is the Euler-Lagrange form of $\lambda$ (i.e.
$\mathcal{E}^{ij}=0$ are the Euler-Lagrange equations for
$\lambda$) and $\theta\in\Omega^{n-1,1}(J\mathcal{M}_{M})$ is the
symplectic potential. Another example is equation (\ref{Euler})
bellow.

If $\lambda$ is $\mathrm{Diff}M$-invariant we have
$0=L_{X_J}\lambda=\iota_{V_{X}}\delta\lambda+d\iota_{H_{X}}\lambda=\iota_{V_{X}}
\mathcal{E}+d\iota_{V_{X}}\theta+d\iota_{H_{X}}\lambda$, and hence
the conserved current can be defined by
$J(X)=\iota_{V_{X}}\theta+\iota_{H_{X}}\lambda$ as we have
$dJ(X)=-\iota_{V_{X}} \mathcal{E}\approx 0$.

\subsection{Pontryagin forms}

At every point of $J\mathcal{M}_{M}$ the first derivatives can be
used to define the Christoffel symbols
$\Gamma_{jk}^{i}=\frac{1}{2}g^{ia}\left(
g_{ak,j}+g_{aj,k}-g_{jk,a}\right)  $ and a covariant derivative
$\left( D^{\Gamma}A\right) ^{i}=DA^{i}+\Gamma_{jk}^{i}A^{j}dx^{k}$
for $A=A^{i}
\partial/\partial x^{i}\in\Gamma(J\mathcal{M}_{M},TM)$. It can be shown that
in this way we obtain a well defined connection $\Gamma$ (called
the horizontal Levi -Civita connection) and that $\Gamma$ is
invariant under the natural action of $\mathrm{Diff}M$ (see
\cite{natconn} for details).

Let $\Omega\in\Omega^{2}(J\mathcal{M}_{M},\mathrm{End}TM)$ be the
curvature of the connection $\Gamma$. Then locally we have
\begin{eqnarray}
\Omega_{j}^{i}  &  =D\Gamma_{jk}^{i}\wedge
dx^{k}+\Gamma_{as}^{i}\Gamma
_{jr}^{a}dx^{s}\wedge dx^{r}\nonumber\\
&  =\delta\Gamma_{jk}^{i}\wedge dx^{k}+\frac{1}{2}R_{jsr}^{i}
dx^{s}\wedge dx^{r} \label{Omegahor}
\end{eqnarray}

We write this equation simply by $\Omega=\delta\Gamma+R$. Note
that this decomposition corresponds to that in (\ref{des}). By
applying the first Pontryagin polynomial to $\Omega$ we obtain the
first Pontryagin form
$p_{1}(\Omega)=-\frac{1}{8\pi^{2}}\mathrm{tr}(\Omega\wedge\Omega
)\in\Omega^{4}(J\mathcal{M}_{M})$. For simplicity we set
$P=-8\pi^{2} p_{1}(\Omega)$. In dimension $3$ if we consider the
components of this form, using formula (\ref{Omegahor}) we obtain
$P=P_{1}+P_{2}$ with $P_{1} =2\mathrm{tr}(\delta\Gamma\wedge
R)\in\Omega^{3,1}(J\mathcal{M}_{M})$,
$P_{2}=\mathrm{tr}(\delta\Gamma\wedge\delta\Gamma)\in\Omega
^{2,2}(J\mathcal{M}_{M})$.

Equation (\ref{EP1}) expressed in terms of forms in the jet bundle
gives
\begin{equation}
P_1=2\mathrm{tr}(\delta\Gamma\wedge
R)=\mathcal{C}-d\eta\label{Euler}
\end{equation}
where
\begin{eqnarray*}
\mathcal{C}  &  =C^{ab}\delta g_{ab}\wedge\mathrm{vol}\in
F^{1}\subset
\Omega^{3,1}(J\mathcal{M}_{M})\\
\eta &  =-2R_{b}^{i}\delta g_{ia}\wedge dx^{a}\wedge
dx^{b}\in\Omega ^{2,1}(J\mathcal{M}_{M})
\end{eqnarray*}
In this case both $\mathcal{C}$\ and $\eta$ are
$\mathrm{Diff}M$-invariant.

\subsection{Equivariant Pontryagin forms}

We recall the definition of equivariant Pontryagin forms given in
\cite{WP}. They are used in \cite{anomalies} to study the problem
of local gravitational anomaly cancellation, and in \cite{WP,GeoD}
are shown to be related to symplectic structures and moment maps
in the space of metrics in dimensions $4k-2$. In this paper we
show that they are related to conserved currents and black hole
entropy of Chern-Simons terms in dimension $3$.

We recall that when a connection is invariant under the action of
a group $G$, in addition to the ordinary characteristic classes,
we can consider the corresponding $G$-equivariant characteristics
classes, which are closed under the Cartan differential (see
\cite{BV1,BGV}). In our case the connection $\Gamma$ is invariant
under the action of $\mathrm{Diff}M$ and we can consider the
$\mathrm{Diff}M$-equivariant Pontryagin forms.

The construction of equivariant Pontryagin forms is based on the
following equation (see \cite{WP}), which can be obtained directly
from equations (\ref{HX}), (\ref{VX}) and (\ref{Omegahor})
\begin{equation}
\iota_{X_{J}}\Omega=-D^{\Gamma}(\nabla X), \label{equicurv}
\end{equation}
where $\nabla X_{b}^{a}=\partial_{b}X^{a}+\Gamma_{bc}^{a}X^{c}$.
In the decomposition (\ref{des}) this equation becomes
\begin{eqnarray}
\iota_{V_{X}}\delta\Gamma+\iota_{H_{X}}R    =-d^{\Gamma}(\nabla
X),\label{equicurv1}\\
\iota_{H_{X}}\delta\Gamma   =-\delta(\nabla X). \label{equicurv2}
\end{eqnarray}

The first equivariant Pontryagin form is defined by (see
\cite{WP})
\[
-\frac{1}{8\pi^{2}}\mathrm{tr}(\Omega-\nabla
X)^{2}=-\frac{1}{8\pi^{2}
}\mathrm{tr}(\Omega\wedge\Omega)+\frac{1}{4\pi^{2}}\mathrm{tr}
(\nabla X\cdot\Omega)-\frac{1}{8\pi^{2}}\mathrm{tr}(\nabla X^{2}),
\]
for $X\in\mathfrak{X}(M)$, and is closed under the Cartan
differential $D_{C}=D-\iota_{X_{J}}$ by virtue of equation
(\ref{equicurv}). In particular this implies that we have
$\iota_{X_{J}}P=D\left( \beta(X)\right)  $, where
$\beta(X)=2\mathrm{tr}(\nabla X\cdot\Omega)$.

A map
$\mu\colon\mathfrak{X}(M)\rightarrow\Omega^{k}(J\mathcal{M}_{M})$
is $\mathrm{Diff}M$-equivariant if $L_{Y_{J}}(\mu(X))=\mu([Y,X])$
for every $X,Y\in\mathfrak{X}(M)$. For example, if
$\alpha\in\Omega^{k+1}(J\mathcal{M} _{M})$ is
$\mathrm{Diff}M$-invariant, then the map $X\mapsto\iota_{X_{J}
}\alpha$ is equivariant. One of the properties of equivariant
Pontryagin forms is that the form
$\beta\colon\mathfrak{X}(M)\rightarrow\Omega^{2}
(J\mathcal{M}_{M})$ is $\mathrm{Diff}M$-equivariant.

According to decomposition (\ref{des}) we have $\beta(X)=\beta
_{0}(X)+\beta_{1}(X)$ , where
\begin{eqnarray}
\beta_{0}(X)  &  =2\mathrm{tr}(\nabla X\cdot R)\in\Omega^{2,0}
(J\mathcal{M}_{M}),\label{Ebeta}\\
\beta_{1}(X)  &  =2\mathrm{tr}(\nabla X\cdot\delta\Gamma)\in\Omega
^{1,1}(J\mathcal{M}_{M}). \label{Ebeta2}
\end{eqnarray}

We show below that $\beta_{0}$ appears on the computation of the
conserved current and $\beta_{1}$ appears on the computation of
the hamiltonian current and the black hole entropy.

Under the decomposition (\ref{des}) into horizontal and vertical
terms equation $\iota_{X_J}P=D\left( \beta(X)\right) $ becomes
\begin{eqnarray}
\iota_{V_{X}}P_{1}   =d\beta_{0}(X),\label{ixp1}\\
\iota_{H_{X}}P_{1}+\iota_{V_{X}}P_{2} =\delta\beta_{0}(X)+d\beta
_{1}(X),\label{ixp2}\\
\iota_{H_{X}}P_{2}    =\delta\beta_{1}(X). \label{ixp3}
\end{eqnarray}

\subsection{Chern-Simons forms}\label{CSF}

As commented before, the expression (\ref{csl})\ for the
Chern-Simons lagrangian is only valid locally. However, we show
that the Cotton tensor admits a global lagrangian which can be
constructed by fixing a metric.

Let $\overline{g}\in\mathrm{Met}M$ be a fixed metric on $M$, let
$\overline{\Gamma}$ be its Levi-Civita connection and
$\overline{R}$ its curvature. Then by applying the usual
transgression formula we obtain $\mathrm{tr}\left(
\Omega^{2}\right)  -\mathrm{tr}(\overline
{R}^{2})=D\overline{CS}$, where
\[
\overline{CS}=2\mathrm{tr}\left(
\overline{a}\wedge\overline{R}\right) +\mathrm{tr}\left(
\overline{a}\wedge D^{\overline{\Gamma}}\overline {a}\right)
+\frac{2}{3}\mathrm{tr}\left(  \overline{a}^{3}\right)  .
\]
and $\overline{a}=\Gamma-\overline{\Gamma}$. However, $\mathrm{tr}
(\overline{R}^{2})=0$ because it is a horizontal $4$-form, and
hence $P=\mathrm{tr}\left(  \Omega^{2}\right) =D\overline{CS}$.

Accordingly with the decomposition (\ref{des})\ we have $\overline
{CS}=\overline{CS}_{0}+\overline{CS}_{1}$ where the second term is
$\overline{CS}_{1}=\mathrm{tr}(\overline{a}\wedge\delta\Gamma)\in
\Omega^{2,1}(J\mathcal{M}_{M})$, and the first term
$\overline{\lambda}
_{CS}=\overline{CS}_{0}\in\Omega^{3,0}(J\mathcal{M}_{M})$ is a
lagrangian density
\[
\overline{\lambda}_{CS}=2\mathrm{tr}\left( \overline{a}\wedge
\overline{R}\right)  +\mathrm{tr}\left( \overline{a}\wedge
d^{\overline{\Gamma}}\overline{a}\right)
+\frac{2}{3}\mathrm{tr}\left( \overline{a}^{3}\right)
\]
If in a local chart we choose $\overline{g}$ a constant metric
then $\overline{\lambda}_{CS}=\mathrm{tr}\left( \Gamma\wedge
d\Gamma +\frac{2}{3}\Gamma^{3}\right)  $ is the usual expression
of the Chern-Simons lagrangian. Hence $\overline{\lambda}_{CS}$ is
a globally well defined lagrangian density with generalizes
(\ref{csl}).

The equation $P=D\overline{CS}$ expressed in terms of the
decomposition (\ref{des}) gives
\begin{eqnarray}
P_{1}  &  =\delta\overline{CS}_{0}+d\overline{CS}_{1},\label{PdC1}\\
P_{2}  &  =\delta\overline{CS}_{1}. \label{PdC2}
\end{eqnarray}

Obviously $\overline{CS}$ is not $\mathrm{Diff}M$-invariant as it
depends on the metric $\overline{g}$. In fact we have
\[
L_{X_{J}}\overline{CS}=\iota_{X_{J}}D\overline{CS}+D\iota_{X_{J}}\overline
{CS}=\iota_{X_{J}}P+D\iota_{X_{J}}\overline{CS}=D(\beta(X)+\iota_{X_{J}
}\overline{CS}).
\]
In the decomposition (\ref{des}) this equation becomes
\begin{eqnarray}
L_{X_{J}}\overline{\lambda}_{CS}  &  =d(\beta_{0}(X)+\iota_{V_{X}}
\overline{CS}_{1}+\iota_{H_{X}}\overline{\lambda}_{CS}).\label{LCS0}\\
L_{X_{J}}\overline{CS}_{1}  &  =\delta(\beta_{0}(X)+\iota_{H_{X}}
\overline{\lambda}_{CS}+\iota_{V_{X}}\overline{CS}_{1})+d(\beta_{1}
(X)+\iota_{H_{X}}\overline{CS}_{1}) \label{LCS1}
\end{eqnarray}

By equation (\ref{Euler}) we have $P_{1}=\mathcal{C}-d\eta$. Using
this equation and equation (\ref{PdC1}) we obtain the first
variational formula for $\overline{\lambda}_{CS}$
\begin{equation}
\delta\overline{\lambda}_{CS}
=P_{1}-d\overline{CS}_{1}=\mathcal{C}-d\left(
\eta+\overline{CS}_{1}\right)  =\mathcal{C}-d\overline{\theta},
\label{FVF}
\end{equation}
where we define the symplectic potential by
$\overline{\theta}=\eta +\overline{CS}_{1}$ (this equation
generalizes formula (\ref{dlambda})). In particular the Cotton
form $\mathcal{C}$ is the Euler-Lagrange form of the Chern-Simons
lagrangian. Hence $\overline{\lambda}_{CS}$ is a globally defined
lagrangian density, whose Euler-Lagrange operator is the Cotton
tensor.

\section{Conserved current for the Cotton tensor\label{current}}

The construction of the conserved current for the Chern-Simons
term is based on equations (\ref{Euler}) and (\ref{Ebeta}). By
combining these equations we obtain
\[
d\beta_{0}=\iota_{V_{X}}P_{1}=2\delta_{X}g_{ab}C^{ab}\mathrm{vol}
+d\iota_{V_{X}}\eta.
\]

If we define
\[
J_{CS}(X)=\alpha(\iota_{V_{X}}\eta-\beta_{0}(X))
\]
then we have
$dJ_{CS}(X)=-2\alpha\delta_{X}g_{ab}C^{ab}\mathrm{vol}\approx0$
and hence $J_{CS}(X)$ is a conserved current. Moreover, the
invariance of $\eta$ and the $\mathrm{Diff}M$-equivariance of
$\beta_{0}$ imply that the map
$J_{CS}\colon\mathfrak{X}(M)\rightarrow\Omega^{2,0}(J\mathcal{M}_{M})$
is $\mathrm{Diff}M$-equivariant. An alternative way to obtain the
same current is by combining equations (\ref{LCS0}) and
(\ref{FVF}).

Next we compute the explicit expression of the conserved current
using the results of the preceding sections. We have
\begin{equation}
\iota_{V_{X}}\eta=-2R_{c}^{i}\left(
g_{ja}\nabla_{i}X^{j}+g_{ji}\nabla _{a}X^{j}\right)  dx^{a}\wedge
dx^{c} \label{iveta}
\end{equation}

Moreover, using the formula (\ref{expR}) in the expression
(\ref{Ebeta}) of $\beta_{0}(X)$ we obtain
\begin{equation}
\beta_{0}(X)=2\left(  \left(
g_{ji}\nabla_{a}X^{j}-g_{ja}\nabla_{i} X^{j}\right)
R_{c}^{i}-\frac{1}{2}R\nabla_{a}X^{j}g_{jc}\right) dx^{a}\wedge
dx^{c} \label{expbeta}
\end{equation}

Finally using (\ref{iveta}) and (\ref{expbeta}) we obtain the
expression for the conserved current $J_{CS}(X)$ given by
\begin{equation}
J_{CS}(X)= -4\alpha
\nabla_{a}X^{j}(R_{jb}-\frac{1}{4}Rg_{jb})dx^{a}\wedge dx^{b}
\label{JF}
\end{equation}

It is well known that the conserved currents corresponding to
$\mathrm{Diff}M$ invariance are globally exact on shell (e.g. see
\cite{Wald}), i.e. we have $J_{CS}(X)\approx dQ_{CS}(X)$ for every
$X\in\mathfrak{X}(M)$ where $Q_{CS}(X)$ is called\ the Noether
charge. In our case, if we define the Noether charge by

\begin{equation}
Q_{CS}(X)=-4\alpha X^{i}(R_{ij}-\frac{1}{4}Rg_{ij})dx^{j}
\label{expQ}
\end{equation}
then, a direct computations shows that we have
\begin{equation}
dQ_{CS}(X)=J_{CS}(X)-4\alpha
g_{ka}C^{ia}X^{k}\mathrm{vol}_{i}\approx J_{CS}(X), \label{dQ}
\end{equation}
where $\mathrm{vol}_{i}=\iota_{\partial_{i}}\mathrm{vol}$.
Moreover, the map
$Q\colon\mathfrak{X}(M)\rightarrow\Omega^{1,0}(J\mathcal{M}_{M})$
is $\mathrm{Diff}M$-equivariant. The tensor
$S_{ij}=R_{ij}-\frac{R}{4}g_{ij}$ which appears in the expression
of the Noether potential is called the Schouten tensor.

\section{Black hole entropy in the presence of Chern-Simons
terms\label{entropy}}

In this section we follow Wald's Noether charge approach to
compute the black hole entropy corresponding to the Chern-Simons
term in $3$ dimensions by using the results of the previous
sections. Our approach is similar to that in \cite{Tach} but we
use the geometrical constructions of the previous sections. As it
is shown below, in the computation of the black hole entropy the
Noether charge disappears, but it appears an additional term
related to the form $\beta_{1}(X)$ given by the equivariant
Pontryagin form.

First we recall the basic ideas of Wald's Noether charge method
for computing black hole entropy for a $\mathrm{Diff}M$-invariant
lagrangian. Let $\lambda\in\Omega^{n,0}(J\mathcal{M}_{M})$ be a
$\mathrm{Diff}M$ invariant lagrangian density and suppose that we
have the first variational formula
$\delta\lambda=\mathcal{E}-d\theta$, where $\mathcal{E}\in\Omega
^{n,1}(J\mathcal{M}_{M})$ is the Euler-Lagrange form of $\lambda$
and $\theta\in\Omega^{n-1,1}(J\mathcal{M}_{M})$ is the symplectic
potential, and both of them are $\mathrm{Diff}M$-invariant. The
conserved current is given by
$J(X)=\iota_{V_{X}}\theta+\iota_{H_{X}}\lambda$ . Moreover, we
have $J(X)\approx dQ(X)$ where $Q(X)$ is the Noether charge.

The form $\omega=\delta\theta\in\Omega^{n-1,2} (J\mathcal{M}_{M})$
determines a presymplectic structure on the space of extremals by
setting
\[
\sigma_{g}(Y,Z)=\int_{\Xi}jg^{\ast}(\iota_{Z_{J}}\iota_{Y_{J}}\omega)
\]
where $\Xi$ is a Cauchy hypersurface, $g$ is an extremal metric,
and $Y,Z$ are Jacobi fields on $g$ (i.e. variations of the metric
satisfying the linearized equations $L_{Y} \mathcal{E}|g=0$) and
if locally we have $Y=y_{ij}dx^{i}dx^{j}$ then $Y_{J}$ is the
vector field on $J\mathcal{M}_{M}$ given by
$Y_{J}=\sum_{ij,I}\frac{\partial^{|I|}y_{ij}}{\partial
x^{I}}\frac{\partial}{\partial g_{ij,I}}$.

Using the invariance of $\theta$ we obtain
\begin{equation}
\iota_{V_{X}}\omega\simeq d(\delta
Q(X)-\iota_{H_{X}}\theta),\label{ff}
\end{equation}
where $\simeq$ means modulo terms that vanish when the form is
contracted with a Jacobi field and evaluated in an extremal
metric. Hence we have
\begin{equation}
\iota_{\delta_{X}g}\sigma=\int_{\Xi}d(\delta
Q(X)-\iota_{H_{X}}\theta
)=\delta\int_{\partial\Xi}Q(X)-\int_{\partial\Xi}\iota_{H_{X}}\theta\label{fg}
\end{equation}
If $\partial\Xi$ is an asymptotic $(n-2)$-sphere at infinity
$\Sigma_\infty$ and $\iota_{H_{X}}\theta=\delta B(X)$ for certain
$B(X)\in\Omega^{n-2,0} (J\mathcal{M}_{M})$ then
$H(X)=\int_{\Sigma_\infty }(Q(X)-B(X))$ can be considered as the
Hamiltonian function corresponding to the vector field $X$, as it
satisfies Hamilton's equation $\iota_{\delta_{X} g}\sigma=\delta
H(X)$.

Now let us suppose that we have a stationary black hole spacetime
with a bifurcate Killing horizon $\Sigma$ generated by $\xi$, and
let $\Xi$ be an asymptotically flat surface having $\Sigma$ as its
only interior boundary. As $\xi$ is a Killing vector field we have
$\delta_{\xi}g=0$ and the right hand side of equation (\ref{fg})
vanishes. If $\xi=\partial_t+\Omega\partial_\phi$ where
$\partial_t$ is the generator of the global time translation,
$\Omega$ the angular velocity of the horizon and $\partial_\phi$
the angular rotation then we obtain
\[
\delta\int_{\Sigma}Q(\xi)= \delta\int_{\Sigma_\infty}Q(\partial_t)
+\Omega\delta \int_{\Sigma_\infty}Q(\partial_\phi)-
\int_{\Sigma_\infty}\iota_{H_{\partial_t}}\theta
\]
where $\Sigma_\infty$ is the asymptotic infinity of the Cauchy
surface, and we have used that $\xi|_{\Sigma}=0$ and hence
$\iota_{H_{\xi}}\theta|_{\Sigma}=0$, and that
$\iota_{H_{\partial_\phi}}\theta|_{\Sigma_{\infty}}=0$ because
$\partial_{\phi}$ is tangent to $\Sigma_{\infty}$.

By defining the energy by
$\mathcal{E}=\int_{\Sigma_\infty}(Q(\partial_t)-B(\partial_t))$,
the angular momentum by
$\mathcal{J}=-\int_{\Sigma_\infty}Q(\partial_\phi)$ and the
entropy by $S=\frac{2\pi} {\kappa}\int_{\Sigma}Q(\xi)$ we obtain
the first law of black hole thermodynamics $\kappa\delta S=\delta
\mathcal{E}-\Omega\delta \mathcal{J}$ where $\kappa$ is the
surface gravity of the black hole characterized by
$\nabla\xi|_{\Sigma} =\kappa\epsilon$ and $\epsilon$ is the
binormal to $\Sigma$.

For a $\mathrm{Diff}M$-invariant lagrangian Wald's formula gives a
general expression (see \cite{Wald,IW,Jacobson})
\[
S=-2\pi\int_{\Sigma}\frac{\delta L}{\delta
R_{abcd}}\epsilon_{ab}\varepsilon_{cdi_1\dots
i_{d-2}}dx^{i_1}\dots dx^{i_{d-2}}
\]
where $\lambda=L\mathrm{vol}$ and $L$ is considered as a function
of the curvature tensor $R$ and its covariant derivatives.

In $3D$ topologically massive gravity the lagrangian includes a
Chern-Simons term which is not $\mathrm{Diff}M$-invariant, and
hence Wald's formula cannot be applied in this case. We follow
Wald's Noether charge approach to obtain a formula for the entropy
and we show that our result coincides with that obtained in
\cite{Tach}.

For the Chern-Simons lagragian we have the first variational
formula (\ref{FVF}). Hence in our case the presymplectic structure
on the space of solutions $\sigma$ is determined by the form
$\alpha\omega$, where
\[
\omega=\delta\overline{\theta}=\delta\eta+\delta\overline{CS}_{1}=\delta
\eta+P_{2}.
\]

Note that although $\overline{\theta}$ depends on the metric
$\overline{g}$ and hence is not $\mathrm{Diff}M$-invariant, the
form $\omega$ is $\mathrm{Diff}M$-invariant. As $\eta$ is
$\mathrm{Diff}M$-invariant we have
\[
L_{X_{J}}\eta=\iota_{H_{X}}d\eta+d\iota_{H_{X}}\eta+\delta\iota_{V_{X}}
\eta+\iota_{V_{X}}\delta\eta=0.
\]

Using this equation we obtain
\begin{equation}
\iota_{V_{X}}\omega=\iota_{V_{X}}\delta\eta+\iota_{V_{X}}P_{2}=-\iota_{H_{X}
}d\eta-d\iota_{H_{X}}\eta-\delta\iota_{V_{X}}\eta+\iota_{V_{X}}P_{2}
\label{iomega}
\end{equation}
By equations (\ref{Euler}) and (\ref{ixp2}) we have
\[
\iota_{H_{X}}d\eta=\iota_{H_{X}}\mathcal{C}-\iota_{H_{X}}P_{1}\simeq
-\delta\beta_{0}(X)-d\beta_{1}(X)+\iota_{V_{X}}P_{2}
\]
Moreover, by equation (\ref{dQ})\ we have
\[
 \iota_{V_{X}}\eta-\beta
_{0}(X)=J_{CS}(X)=dQ_{CS}(X)+2g_{ka}\mathcal{C}^{ia}X^{k}\mathrm{vol}
_{i}\simeq dQ_{CS}(X).
\]
 Replacing this equations on (\ref{iomega})
we obtain
\begin{equation}
\iota_{V_{X}}\omega\simeq d\left(  \delta
Q_{CS}(X)+\beta_{1}(X)-\iota_{H_{X}} \eta\right)  \label{fnf}
\end{equation}

Finally, using equation (\ref{equicurv2}) we have
\[
\beta_{1}(X)=2\mathrm{tr}(\nabla X\cdot\delta\Gamma)=\delta\left(2
\mathrm{tr}(\nabla X\cdot\overline
{a})-\mathrm{tr}(\iota_{H_{X}}\overline{a}\cdot\overline
{a})\right)
+\iota_{H_{X}}\mathrm{tr}(\overline{a}\wedge\delta\Gamma)
 \]
By replacing the last equation on (\ref{fnf}) we obtain the
analogous of equation (\ref{ff}) for the Chern-Simons lagrangian

\begin{equation}
\iota_{V_{X}}\omega\simeq d\left(  \delta
q_{CS}(X)+\iota_{H_{X}}\nu\right)  \label{fn}.
 \end{equation}
where we define
\begin{eqnarray*}
q_{CS}(X)   =Q_{CS}(X)+2\mathrm{tr}(\nabla
X\cdot\overline{a})-\mathrm{tr}(\iota_{H_{X}
}\overline{a}\cdot\overline{a}),\\
\nu   =\mathrm{tr}(\overline{a}\wedge\delta\Gamma)-\eta
\end{eqnarray*}

Hence for the Chern-Simons lagrangian the hamiltonian current
$q_{CS}$ plays the same role as $Q$ for $\mathrm{Diff}M$-invariant
lagrangians in the computation of black hole entropy.

Choosing $\overline{g}$ a constant metric we obtain
$q(X)=Q(X)+\mathrm{tr}(2\nabla
X\cdot\Gamma)-\mathrm{tr}(\iota_{H_{X}}\Gamma \cdot\Gamma).$\ In
local coordinates we have
\[
q_{CS}(X)=\left(
-4S_{ir}X^{i}dx^{r}+2\partial_{j}X^{i}\Gamma_{ir}^{j}+\Gamma
_{jk}^{i}\Gamma_{ir}^{j}X^{k}\right)  dx^{r}
\]
where $S_{ij}$ are the components of the Schouten tensor.

This expression is similar to the current defined in \cite{MO},
given by
\[
q_{2}(X)=\left( -4S_{ir}X^{i}+\Gamma_{jk}^{i}\Gamma_{ir}^{j}
X^{k}\right) dx^{r}
\]
Note that the difference between $q_{CS}$ and $q_2$ are the terms
containing derivatives of $X$, and hence for a constant vector
both expressions coincide. For example in \cite{MO} $q_2(X)$ is
used to compute the contribution of the Chern-Simons term to the
energy (or mass) and angular momentum for BTZ black holes,
Log-gravity and Warped AdS$_3$ black holes by setting
$\mathcal{E}_{CS}=\alpha\int_{\Sigma_{\infty}}q_2(\partial_{t})$
and
$\mathcal{J}_{CS}=\alpha\int_{\Sigma_{\infty}}q_2(\partial_{\phi
})$. If the vector fields $\partial_{t}$ and $\partial_{\phi}$ are
constant we have $\mathcal{E}_{CS}=
\alpha\int_{\Sigma_{\infty}}q_{CS}(\partial_{t})$ and
$\mathcal{J}_{CS}=\alpha\int_{\Sigma_{\infty}}q_{CS}(\partial_{\phi})$.

Moreover, the current $q_{CS}$ also gives the correct result for
the entropy. If $\xi=\partial_{t}+\Omega\partial_{\phi}$, over
$\Sigma$ we have $\xi|_{\Sigma }=0$ and hence
$\iota_{H_{\xi}}\nu=0$\ and $q(\xi)|_{\Sigma}
=2\mathrm{tr}(\nabla\xi\cdot\Gamma)|_{\Sigma}=2\kappa\mathrm{tr}
(\epsilon\cdot\Gamma)$. Hence we can define the contribution of
the Chern-Simons term to the black hole entropy by
\[
S_{CS}=\frac{2\pi\alpha}{\kappa}\int_{\Sigma}q(\xi)=4\pi\alpha\int_{\Sigma
}\mathrm{tr}(\epsilon\cdot\Gamma).
\]
This expression coincides with the result obtained in \cite[\S
3.1]{Tach}. In \cite{Tach}\ this expression is computed for the
BTZ black hole and shown to coincide with the results obtained by
other methods in \cite{S,KL1,KL2,SS,Cho,Park}.

As commented in the Introduction, most of our geometrical
constructions can be extended to higher dimensions. However, the
analogous of equation (\ref{fn}) needed to define the black hole
entropy is not so simple in dimensions greater than $3$.

\end{document}